\documentclass[12pt]{article}
\usepackage{epsfig}
\usepackage{graphicx}

\def\be{\begin{equation}}
\def\ee{\end{equation}}
\def\l{\label}

\begin{document}

\title{Quasicanonical Gibbs distribution and Tsallis nonextensive statistics}
\author{A.K. Aringazin and M.I. Mazhitov}

\author{A.K. Aringazin and M.I. Mazhitov\\
Department of Theoretical Physics, \\Institute for Basic Research,
Eurasian National University,\\ Astana 473021 Kazakstan}


\date{17 April 2002}

\maketitle

\begin{abstract}
We derive and study quasicanonical Gibbs distribution function
which is characterized by the thermostat with finite number of
particles (quasithermostat). We show that this naturally leads to
Tsallis nonextensive statistics and thermodynamics, with Tsallis
parameter $q$ is found to be related to the number of particles in
the quasithermostat. We show that the chi-square distribution of
fluctuating temperature used recently by Beck can be partially
understood in terms of normal random momenta of particles in the
quasithermostat. Also, we discuss on the importance of the time
scale hierarchy and fluctuating probability distribution functions
in understanding of Tsallis distribution, within the framework of
kinetics of dilute gas and weakly inhomogeneous systems.
\end{abstract}


\section{Introduction}

Nonextensive statistical mechanics introduced by Tsallis
\cite{Tsallis} has been developed and applied by many authors
\cite{Tsallis2}-\cite{Wilk}; see \cite{Tsallis3} for a recent
review. Particularly, various aspects of the nonextensive
statistics and thermodynamics have been studied by Abe and
Rajagopal \cite{Abe}. Derivation of power law canonical
distributions from first principles of statistical mechanics has
been analyzed by Almeida \cite{Almeida}; see also \cite{Plastino}.

The entropic parameter $q$ of Tsallis statistics is associated to
the measure of nonextensivity, with the usual extensivity
recovered at $q=1$. In Tsallis formalism, $q$ appears in the
definition of entropy which generalizes the usual
Boltzmann-Gibbs-Shannon entropy.

In this paper we show that a nonextensivity characterized by a
parameter naturally arises from the consideration of the
microcanonical Gibbs distribution. Namely, in Sec.~\ref{Sec:Quasi}
we consider a subsystem weakly interacting with a system
containing big but {\sl finite} number $M$ of particles which we
call quasithermostat. It appears that the resulting equilibrium
distribution function and the associated entropy
(Sec.~\ref{Sec:Entropy}) reproduce exactly those of Tsallis
statistics, with $q-1 \sim 1/M$. In the (thermodynamic) limit, $M
\to \infty$, the usual canonical Gibbs distribution is recovered.

Recently, Beck \cite{Beck} have studied application of Tsallis
formalism to turbulent flows and achieved a very good agreement
with experimental measurements. Particularly, it is worthwhile to
emphasize that no fitting parameters have been used by Beck to
reproduce histogram of the acceleration of a test particle
advected by the turbulent flow, while some other authors have used
four free parameters in some {\it ad hoc} modified exponential
function, for this purpose (see \cite{Beck} for details). The main
idea underlying Tsallis statistical mechanics approach to
turbulence is to introduce fluctuation of temperature or
fluctuation of energy dissipation \cite{Beck2, Wilk} described by
$\chi^2$ distribution of order $m$.

In this paper we show that the relevance of $\chi^2$ distribution
can be understood in terms of normal random momenta of particles
in the quasithermostat, with $m=3M$ representing the total number
of random variables (Sec.~\ref{Sec:Fluctuations}).

Some aspects of fluctuations and time scale hierarchy in dilute
gas and weak inhomogeneous systems are presented in
Sec.~\ref{Sec:Hierarchy}.

In this paper we show the importance of the time scale hierarchy
and fluctuating probability distribution functions in
understanding Tsallis distribution, which is viewed as the average
value of fluctuating canonical Gibbs distribution over the random
parameter $\tilde\beta$.

\section{Quasicanonical Gibbs distribution}\label{Sec:Quasi}

We start with the classical microcanonical Gibbs distribution,
$f_L(x,a,E)$, which describes isolated equilibrium system of $L$
classical particles with total energy $E$,
\be\l{microcan}
f_L(x,a,E) = \frac{1}{\Omega(a,E)}\delta(H(x,a)-E),
\ee
where $ \Omega(a,E) = \int \delta(H(x,a) - E)dx$, $H(x,a)$ is
Hamiltonian of the system, $a$ denotes external parameters,
$dx=\prod_{1\leq i \leq L} dx_i$, and $x_i=({\vec p}_i, {\vec
r}_i)$ are phase-space coordinates of the particles.

In the closed system of $L$ particles we select a subsystem of $N
\gg 1$ particles which is surrounded by the subsystem of $L-N=M$
particles. Assuming $M \gg N$, neglecting the interaction energy
between the subsystems, and taking the limit $M \to \infty$, from
the microcanonical distribution (\ref{microcan}) one can derive,
following one of the known procedures, the usual canonical Gibbs
distribution,
\be\l{can}
f_N(x,a,T) = \frac{1}{Z}\exp\left[-\frac{H_N(x,a)}{kT}\right],
\ee
where $Z = \int \exp\left[-{H_N(x,a)}/{kT}\right] dx$ is the
classical statistical integral. Here, $k$ is Boltzmann constant
and $T$ is an absolute temperature of the subsystem of $M =
\infty$ particles (called thermostat); $dx=\prod_{1\leq i \leq N}
dx_i$.

It is worthwhile to note that the canonical Gibbs distribution
comes also as the only physically relevant solution of
supersymmetry condition in the context of topological quantum
field theory approach to Hamiltonian systems~\cite{Aringazin}.

Below, we follow some procedure assuming that the number of
particles in thermostat is finite, $M <\infty$. Such a thermostat
can be referred to as a {\sl quasithermostat}.

The total Hamiltonian is represented by
\be
H(x,y) = H_N(x) + H_M(y) + H_{\mathrm{int}}(x,y),
\ee
where $x$ and $y$ denote phase-space coordinates of particles in
the considered subsystem and in the quasithermostat, respectively,
and $H_{\mathrm{int}}(x,y)$ characterizes interaction between the
subsystem and quasithermostat. Herebelow, we drop dependence on
the external parameters to simplify notation.

Since the total system is closed, it is described by the
microcanonical distribution,
\be
f_{N+M}(x,y,E) = C\delta(H(x,y)-E),
\ee
provided that it is normalized, $ \int f_{N+M}(x,y,E)dx dy =1$;
$dy =\prod_{1\leq i \leq M} dy_i$. Note that the numbers $M$ and
$N$ are additive integrals of motion. To obtain the distribution
function describing the considered subsystem one should integrate
out phase-space coordinates $y$ of the quasithermostat,
\be\l{fN}
f_N(x) = C\int \delta(E - H_N(x) - H_M(y) -
H_{\mathrm{int}}(x,y))dy,
\ee
For big $N$ and weak short-range interactions one can ignore the
interaction energy, $H_{\mathrm{int}}(x,y) \ll H_N(x)$, in the
above expression.
By representing Hamiltonian of the quasithermostat as a sum of the
kinetic and potential energy,
\be
H_M(y) = K_M({\vec p}_1, \dots, {\vec p}_M)+ U_M({\vec r}_1,
\dots, {\vec r}_M),
\ee
where we take for simplicity
\be\l{KMp}
K_M(p)=\sum\limits_{i=1}^{M}\frac{{\vec p}_i^{\ 2}}{2m}
\ee
and $U_M(r) = \sum_{i\not= j} \Phi(|{\vec r}_i - {\vec r}_j|)$ for
point-like particles of mass $m$, we can rewrite (\ref{fN}) as
\be
f_N(x) = C\!\! \int\!\! \delta(E - H_N(x)- U_M(r) -
z^2)\delta(z-{K_M^{1/2}})dz dp dr,
\ee
where new variable $z$ has been used; $dp= \prod_{1\leq i \leq
M}d\vec p_i$, $dr= \prod_{1\leq i \leq M}d\vec r_i$. Performing
the integral over $z$ with the use of
\be
\int \delta\!\!\left(z-K_M^{1/2}(p)\right)dp \simeq z^{3M-1},
\ee
and general property of Dirac delta function,
\be
\int \delta(g(z)) dz = \int
\sum\limits_k\frac{\delta(z-z_k)}{|g'(z_k)|} dz, \quad g(z_k)=0,
\ee
we get, up to normalization constant,
\be\l{fNx}
f_N(x) = C \int \left(E - H_N(x) - U_M(r)\right)^{\frac{3M}{2}-1}
dr.
\ee
We rearrange this as follows
$$
f_N(x) = C \int \left(1 - \frac{H(x)}{E-
U_M(r)}\right)^{\frac{3M}{2}-1} \times
$$
\be\l{fNxEU}
\times \left(E- U_M(r)\right)^{\frac{3M}{2}-1} dr,
\ee
where we have denoted, for brevity, $H(x) \equiv H_N(x)$.

Denoting the kinetic energy per particle by
\be\l{betakin}
\frac{3}{2\beta} = \frac{K_M(p)}{M},
\ee
which tends to constant for big $M$,
we obtain that for $M \gg N$,
\be\l{EU}
E - U_M(r) \simeq \frac{3M}{2\beta},
\ee
which does not depend on $r$.

Inserting (\ref{EU}) into (\ref{fNxEU})  and performing the
integral over $r$ we obtain, up to normalization constant,
\be\l{fNM}
f_N(x) = C \left(1 - \frac{2}{3M}\beta
H(x)\right)^{\frac{3M}{2}-1}.
\ee

We note that to obtain the canonical Gibbs distribution
(\ref{can}) it remains to take the (thermodynamic) limit,
\be\l{limit}
\lim_{M\to\infty}\left(1 - \frac{2}{3M}\beta
H(x)\right)^{\frac{3M}{2}-1} = \exp\left[ -\beta H(x)\right]
\ee
and identify $\beta^{-1} = kT$.

However, we are interested specifically in the case when $M$ is
finite so that we do not take the above (thermodynamic) limit, and
turn to analysis of the distribution function (\ref{fNM}).

For $M \gg 1$ we make the approximation $3M/2-1 \simeq 3M/2$.
Then, denoting
\be\l{qM}
q-1 = \frac{2}{3M} \geq 0,
\ee
we have from (\ref{fNM})
\begin{eqnarray}\l{fNq}
f_N(x, a) = \frac{1}{Z_q} \left(1 - (q-1)\beta
H(x,a)\right)^{\frac{1}{q-1}} \nonumber\\ \equiv \frac{1}{Z_q}
\exp_{q-1}[-\beta H(x,a)].
\end{eqnarray}
Here, we have introduced $Z_q = \int \exp_q[-\beta H(x,a)]dx$ due
to the normalization condition, $\int f_N dx =1$, $q$-exponential
is defined by $\exp_s[z]=(1+sz)^{1/s}$, $s=q-1$, and we have
restored dependence on the external parameters $a$. In view of the
fact that the number $M$ is taken to be big but finite, i.e.
$q\simeq 1$, the normalized distribution function (\ref{fNq}) can
be referred to as the {\sl quasicanonical Gibbs distribution}.

The obtained distribution function (\ref{fNq}) strongly resembles
the distribution function in Tsallis nonextensive statistics
\cite{Tsallis, Tsallis2, Tsallis3}, where $q$ is called entropic
parameter.

We remind that $q$-exponential, $\exp_{q-1}[z]$, reduces to the
ordinary exponential, $\exp[z]$, at $q\to 1$, so that the
canonical Gibbs distribution is recovered.

\begin{figure}[ht]
\begin{center}
\epsfxsize=0.85\textwidth
\parbox{\epsfxsize}{\epsffile{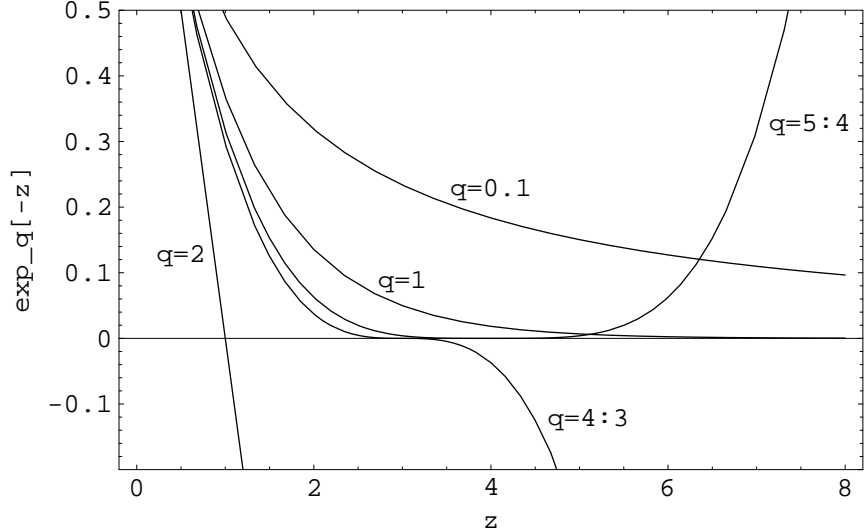}}
\end{center}
 \caption{\label{Fig1} The $q$-exponential
 $\exp_{q-1}[-z]=(1-(q-1)z)^{1/(q-1)}$ at various values of $q$.}
\end{figure}

Typical plots of $\exp_{q-1}[-z]$ at various $q$ are presented in
Fig.~\ref{Fig1}. In the physical context, to avoid the negative
and increasing values of $\exp_{q-1}[-z]$ one must restrict
positive $z$ by a suitable condition (the high-energy cut-off).

Some useful formulas with $q$-exponential are:
\be
d\exp_{q-1}[\pm z]/dz = \pm (\exp_{q-1}[\pm z])^{2-q},
\ee
\be
 \int\exp_{q-1}[\pm z]dz = \pm q^{-1}(\exp_{q-1}[\pm z])^q,
\ee
\be
 \exp_{q-1}[-z] = \left(\exp_{1-q}[z]\right)^{-1},
\ee
\be\l{nonfact}
\exp_{q-1}[z_1]\exp_{q-1}[z_2]=\exp_{q-1}[z_1+z_2+(q-1)z_1z_2],
\ee
the $q$-Poisson integral,
\be
 \int z^n e_{q-1}^{-c z^2}dz =
 z^{n+1}\, {_2}F\!{_1}(\frac{n+1}{2}, \frac{1}{1-q},
 \frac{n+3}{2}, (q-1)cz^2),
\ee
 and
\be
 \int_{0}^{\infty}z^n e_{q-1}^{-c z^2}dz =
\frac{1}{2}(c-cq)^{-(n+1)/2}\times$$
$$\times \Gamma\!\left(\frac{n+1}{2}\right)\Gamma\!\left(\frac{q+n(q-1)+1}{2(1-q)}\right)
 \Gamma^{-1}\!\left(\frac{1}{1-q}\right),
\ee
for $n>-1$, $\frac{q+n(q-1)+1}{2(1-q)}>0$,
$\frac{n+1}{2}-\frac{1}{1-q}>0$, and real $c$, where ${_2}F\!{_1}$
is hypergeometric function and $\Gamma$ is Euler gamma function.

The reflection of the $q$-axis with respect to the point $q=1$,
maps $\exp_{q-1}[z]$ to $\exp_{1-q}[z] = (1+(1-q)z)^{1/(1-q)}$
which is often used in the literature as a definition of the
$q$-exponential.

Also, it is remarkable to note that the slightly generalized
$q$-exponential,
\be
\exp_{{q-1},b}[z]\equiv
(1+(q-1)z)^{1/(q-1)+b},
\ee
can be represented as a power of the $q$-exponential,
\be
(1+(q-1)z)^{c/(q-1)}= (\exp_{q-1}[z])^c,
\ee
for the parameter $c =1+(q-1)b$. For example,
$(1+(q-1)z)^{1/(q-1)+1}= (1+(q-1)z)^{q/(q-1)} =(\exp_{q-1}[z])^q$,
so that for small $q-1$ we have $\exp_{q-1}[z] \simeq
(\exp_{q-1}[z])^q$.

Note that $(\exp_{q-1}[z])^c$ corresponds to the scaling of $z$ in
the $q\to 1$ limit, namely, $(\exp_{q-1}[z])^c \to \exp[cz]$,
while in general $(\exp_{q-1}[z])^c \not= \exp_{q-1}[cz]$.

In contrast to the canonical case, the distribution function
(\ref{fNq}) depends not only on the temperature ($\beta^{-1}$) but
also on the total number of particles in quasithermostat displayed
by the parameter $q$. Also, the property (\ref{nonfact}) of
$q$-exponential leads to a non-factorizable character of the
distribution function (\ref{fNq}).

Using the identity $\exp_{q-1}[z]=\exp\{\ln\exp_{q-1}[z]\}$ and
the new variable $z_q = \ln\exp_{q-1}[z] \equiv
(q-1)^{-1}\ln(1+(q-1)z)$, one can formally rewrite $q$-exponential
in the usual exponential form, $\exp_{q-1}[z]=\exp[z_q]$. This
allows to reformulate any $q$-exponential distribution function,
e.g., (\ref{fNq}), in a factorizable form with respect to $z_q$.
Consequently, one can formally achieve an additive character of
Tsallis formalism with respect to such ''$q$-deformed`` entities
$z_q$.

Almeida \cite{Almeida} has derived Tsallis distribution
(\ref{fNq}) from the first principles of statistical mechanics
using the parametrization
\be\l{capacity}
\frac{d \theta}{dE} = q-1,
\ee
where $\theta \equiv 1/\beta$ is the statistical temperature.
Tsallis parameter $q$ is given a physical interpretation in terms
of heat capacity of the heat bath due to the fact that $\theta$ is
related to the temperature of the heat bath. For an infinite heat
capacity (thermostat), $q=1$, one recovers canonical Gibbs
distribution while for {\sl finite} heat capacity
(quasithermostat), $q\not=1$, one gets power law distributions.

The microcanonical approach meets this viewpoint as it establishes
dependence of $q$ on the number of particles $M$ of the
quasithermostat which evidently characterizes its heat capacity,
${d\theta}/{dE} \sim 1/M$. It is worthwhile to note however that
in general the form of the dependence $\theta(E)$ in the l.h.s. of
Eq.~(\ref{capacity}) seems to be not fixed by first principles of
statistical mechanics. The linear dependence, $\theta(E) =
(q-1)E$, does yield Eq.~(\ref{capacity}) while a quadratic
dependence would require yet another parameter. In contrast, the
used microcanonical approach with the kinetic energy in the form
(\ref{KMp}) suggests only one parameter, $M$, which we relate to
$q$ via Eq.~(\ref{qM}).

\section{The entropy}\label{Sec:Entropy}

The entropy associated to the distribution function $f_N =
f_N(H(x,a),\beta,q)$ defined by (\ref{fNq}) can be readily
obtained, up to a factor and constant term.

First, we identically rewrite Eq.~(\ref{fNq}) as
\be
f_N = \exp_{q-1}[\beta F_q-(\beta-(q-1)F_q)H(x,a)],
\ee
where we have denoted
\be
F_q=\beta^{-1}\ln_{q-1}Z_q^{-1},
\ee
and $q$-logarithm is an inversion of $q$-exponential, $\ln_s[z]
=({z^s-1})/s$, $s=q-1$.

Some useful formulas with $q$-logarithm are:
\be
 \lim_{q\to 1}\ln_{q-1} z = \ln z, \quad
 \ln_{q-1} z^{-1} = -\ln_{1-q}z,
\ee
\be
 \ln_{q-1}z^p = p \ln_{p(q-1)}z, \quad
 z\ln_{q-1}z = z^q \ln_{1-q}z,
\ee
\be
 d\ln_{q-1} z/dz = z^{q-2} = z^{-1}(1+(q-1)\ln_{q-1}z),
\ee
\be
\int\ln_{q-1} z dz = q^{-1}z(\ln_{q-1} z -1),
\ee
and
\be\l{nonadd}
\ln_{q-1} z_1z_2 = \ln_{q-1} z_1 +
\ln_{q-1} z_2 + (q-1)\ln_{q-1} z_1 \ln_{q-1} z_2,
\ee

Then, with the use of the inverse function,

\be
H = \rho(f_N) = (\beta F_q-\ln_{q-1} f_N)/(\beta-(q-1)F_q),
\ee
the normalization condition, $\int f_N dx =1$, the thermodynamic
heat, $\delta Q = \int H(x,a) (\delta f_N) dx$, and the
integrability condition, $dQ = T(\beta)dS_q$ one can obtain (cf.
\cite{Almeida})
\be\l{SqaT}
S_q(a,T) = -\!\!\int\!\! f_N \ln_{q-1} f_N dx \equiv -\!\!\int\!\!
f_N^q
\ln_{1-q} f_N dx.
\ee
Here, the representation
\be
S_q = T^{-1}(\beta)\int_{0}^{f_N} (\rho(z)+ \mathrm{const})\, dz
dx
\ee
with a suitable constant has been used. The integrability factor
is found to be $T(\beta) = (q\beta-q(q-1)F_q)^{-1}$.

With $\beta^{-1} = kT$, we obtain
\be
\delta Q = \left(\frac{q}{kT} - q(q-1)F_q\right)dS_q.
\ee

In the discrete case characterized by microstates with energies
$E_n$, the obtained quasicanonical distribution function
(\ref{fNq}) and the entropy (\ref{SqaT}) are written as
\be\l{fnq}
f_n = \frac{1}{Z_q}\exp_{q-1}[-\beta E_n],
\ee
\be
 Z_q =  \sum\limits_{n=1}^{W}\exp_{q-1}[-\beta E_n],
 \quad \sum\limits_{n=1}^{W}f_n = 1,
\ee
\be\l{qentropy}
S_q = -\sum\limits_{n=1}^{W}f_n \ln_{q-1} f_n \equiv
-\sum\limits_{n=1}^{W}f_n^q \ln_{1-q} f_n
\ee
$$
 \equiv -\frac{\sum_{n=1}^{W}(f_n^q - f_n)}{q-1}
 = -\frac{\sum_{n=1}^{W}f_n^q - 1}{q-1}.
$$

The entropy (\ref{qentropy}) coincides with Tsallis entropy
\cite{Tsallis}, with $q$ treated as an entropic parameter, and
reduces to Boltzmann-Gibbs-Shannon entropy, $S=-\sum f_n \ln f_n$,
at $q \to 1$. It is remarkable to note that either $f_n$ and
$f_n^q$ can be used for the averaging in Eq.~(\ref{qentropy})
provided a suitable choice of $q$.

It is well known that by extremizing the (convex) entropy
(\ref{qentropy}) under suitable constraints \cite{Tsallis3} one
obtains Tsallis distribution (\ref{fnq}).

Due to the property (\ref{nonadd}) of $q$-logarithm the entropy
(\ref{qentropy}) is non-additive,
\be\l{Snonadd}
S_q(A,B) = S_q(A) + S_q(B|A) + (q-1)S_q(A)S_q(B|A),
\ee
where we use the conditional entropy for statistically dependent
systems $A$ and $B$ (e.g., subsystems with long-range
interactions) based on the conditional distribution function
$f(A,B) = f(A|B)f(B)=f(B|A)f(A)$.

In the case $A$ and $B$ are statistically independent in a
conventional sense (e.g., subsystems with short-range
interactions), we have formally $S_q(A|B) = S_q(A)$ and $S_q(B|A)
= S_q(B)$.

We note that the $(q-1)$ term in Eq.~(\ref{Snonadd}) is non-zero
even for such "statistically independent" systems contrary to a
naive expectation that the statistical independence should imply
an additive character of the entropy $S_q$. The origin of this
implication seems to be due to the fact that here the parameter
$q$ does not depend by construction on a scale or correlation
parameter, such as a correlation length, which allows one to
determine a statistical independence of the systems $A$ and $B$ in
a physical context (when the systems $A$ and $B$ are separated by
distances much bigger than the correlation length their evolutions
are physically independent).

Formally, for a {\sl constant} parameter $q\not=1$ the systems
appear to be always statistically dependent on each other. Deeper
analysis is required for a consistent formulation of the
statistical independence within the framework of non-additive
formalism (\ref{Snonadd}).

\section{Normal random momenta and fluctuating statistical
temperature}\label{Sec:Fluctuations}

Recently, Beck \cite{Beck} have studied application of Tsallis
formalism to turbulent flows and indicated that the factor of
Tsallis equilibrium distribution function (\ref{fnq}) can be
obtained by averaging Boltzmann factor of the canonical Gibbs
distribution \cite{Johal},
\be\l{beta}
\int\limits_{0}^{\infty} e^{-\tilde\beta E_n} f(\tilde\beta)
d\tilde\beta = (1-(q-1)\beta E_n)^{\frac{1}{q-1}},
\ee
where $1-q = 2/m$ and
\be\l{chi2}
f(\tilde\beta) =
\frac{1}{\Gamma(\frac{m}{2})}\left(\frac{n}{2\beta}\right)^{\frac{m}{2}}
\tilde\beta^{\frac{m}{2}-1}\exp\left[-\frac{m\tilde\beta}{2\beta}\right]
\ee
is the probability density of the $\chi^2$ distribution of order
$m$ (the distribution of sum of squares of $m$ normal random
variables with mean zero and unit variance), with the average
$\langle\tilde\beta\rangle \equiv\int_{0}^{\infty} \tilde\beta
f(\tilde\beta) d\tilde\beta =
\beta$ and the relative variance $(\langle\tilde\beta^2\rangle
-\langle\tilde\beta\rangle^2)/\langle\tilde\beta\rangle^2 = 2/m$.
This is viewed as the result of fluctuation of temperature or
fluctuation of energy dissipation \cite{Beck2, Wilk} and has been
found of particular relevance for nonequilibrium systems. Beck
have used an assumption that the time scale on which $\tilde\beta$
fluctuates is much bigger than that of kinetic relaxation of the
system. This provides a quasistatic behavior of the system with
respect to variation of $\tilde\beta$.

We see that $\chi^2$ randomization and averaging of the canonical
Gibbs distribution over random $\tilde\beta$ leads to Tsallis
distribution.

It seems that the relevance of $\chi^2$ distribution (\ref{chi2})
to relate Boltzmann and Tsallis factors is not due to a
coincidence. By comparing this approach \cite{Beck, Johal} with
the above obtained results (\ref{qM}) and (\ref{fNq}), we see that
the number of random variables is $m=3M$, which is exactly the
total number of components of three-dimensional momenta $p_i$ of
$M$ particles constituting the quasithermostat. Since the
quasithermostat is assumed to be in a quasi-equilibrium state we
can treat the momenta $\vec p_i$ (microscopic characteristics) to
be normal random variables with mean zero and unit deviation as a
simplest model. Hence, the kinetic energy (\ref{KMp}) is $\chi^2$
distributed due to its definition. We associate the kinetic energy
of particles in quasithermostat to its statistical temperature
$\beta^{-1}$ in accord to Eq.~(\ref{betakin}). Hence,
$\tilde\beta^{-1}$ becomes $\chi^2$ random variable.

This partially clarifies the physical origin of the $\chi^2$
distribution used by Beck in Eq.~(\ref{beta}). However, we note
that in our approach $\chi^2$ distribution arises for the
statistical temperature $\tilde\beta^{-1}$ (not for the inverse
statistical temperature $\tilde\beta$).

Usually, normal (not $\chi^2$) distribution of fluctuating
temperature with the relative variance from Eq.~(\ref{DeltaT}) is
used as a good approximation. We note that the averaging of the
canonical Gibbs distribution over {\sl normally} distributed
$\tilde T$ (i.e., $\tilde\beta^{-1}$) with the mean $T$ and some
deviation $\sigma$ is not analytically tractable, while the
averaging over normally distributed $\tilde\beta$ yields the
analytic result,
\be\l{Gauss}
\int_0^\infty\!\! e^{-\tilde\beta H}g(\tilde\beta)d\beta
 = e^{-\beta H
 + \frac{1}{2}\sigma^2H^2}
 \left(1-\mathrm{erf}(\frac{\sigma^2 H -\beta}{\sqrt{2}\sigma})\right),
\ee
where
\be
g(\tilde\beta) =
\frac{1}{\sigma\sqrt{\pi/2}}\exp[-\frac{(\tilde\beta-\beta)^2}{2\sigma^2}]
\ee
is normal distribution, and erf$(x)$ is the error function. The
resulting expression (\ref{Gauss}) includes one special function
and is less handful as compared with Tsallis distribution
(\ref{beta}).

Since the entropy $S_q$ and mean energy $\bar{H}$ of the
equilibrium system are not additive for $q\not= 1$, the function
$F_q$ is not additive as well. Indeed, from Eqs.~(\ref{fNq}) and
(\ref{SqaT}), we have
\be S_q =
(\beta - (q-1)F_q)\bar H - \beta F_q,
\ee
where $\bar H = \int H f_N dx$. This implies that $\beta$ is not a
universal parameter of the system, in contrast to the case of
additive entropy and mean energy. Indeed, $\beta$ is not the only
parameter characterizing the quasithermostat, with $q$ being the
second temperature parameter measuring the relative variance of
the fluctuating parameter around the value $\beta$ viewed as an
average.

From the viewpoint of canonical Gibbs distribution, accounting for
a finiteness of thermostat made in Sec.~\ref{Sec:Quasi} {\sl
automatically} arises to the averaged $\chi^2$ distributed
fluctuation of the parameter $\tilde\beta$ around the mean $\beta$
with the relative variance $\sim 1/M$. Actually we associate the
fluctuations of temperature with eventual fluctuations of the
energy of a finite thermostat. We conclude that the $\chi^2$
statistics is implicit in Tsallis statistical mechanics.

The spatial scale dependence of the entropic parameter, $q=q(r)$,
introduced by Beck \cite{Beck} can be readily understood since $q$
is related to the number of particles. For bigger spatial scale
the fluctuations become smaller so that $q\to 1$ in the big scale
asymptotics.

To illustrate how the resulting nonadditivity affects the
temperature characteristics, we consider two systems with
Hamiltonians $H_1$ and $H_2$, respectively, at the same $\beta$.
Abe and Rajagopal \cite{Abe2} have elaborated the power-low
distribution based thermodynamics of equilibrium to much extent.
Below, we discuss this issue in the context of fluctuations of the
parameter $\beta$.

For weakly interacting canonical systems we have $\exp[-\beta
H_1]\exp[-\beta H_2] = \exp[-\beta H]$, where $H=H_1+H_2$ is
conserved. As usual, this means that the two equilibrium systems
compose an equilibrium system at the same $\beta$, while at
different $\beta$ there is no equilibrium, with conserved $H$. The
$\chi^2$ randomization, $\beta \to \tilde\beta$, and averaging of
each of the two canonical systems, with the same average $\beta$
and relative variance $q-1$, gives us $\exp_{q-1}[-\beta
H_1]\exp_{q-1}[-\beta H_2] = \exp_{q-1}[-\beta H_{\mathrm{eff}}]$,
where we have denoted $H_{\mathrm{eff}}=H+(q-1)\beta H_1 H_2$. On
the other hand, the averaging of the composite system yields a
different result, $\langle \exp[-\beta(H_1+H_2)]\rangle_{\beta} =
\exp_{q-1}[-\beta H]$.

It is misleading to interpret the above relations as that the
composite system appears to be not at thermal equilibrium despite
the identical $\beta$. A thermal equilibrium of two systems in the
sense of their absolute temperatures is well formulated for a {\sl
non-fluctuating} $\beta$ but not for the average, $\beta$, of {\sl
fluctuating} $\tilde\beta$. The extra $\beta$-term in
$H_{\mathrm{eff}}$ looks like a pure effect of the $\chi^2$
averaging since it is proportional to the relative variance $q-1$.
Indeed, in general $\langle f(\tilde\beta,
H_1)\rangle_{\beta}\langle f(\tilde\beta, H_2)\rangle_{\beta}
\not=\langle f(\tilde\beta, H_1)f(\tilde\beta,
H_2)\rangle_{\beta}$ regardless to factorizability of the function
$f(\tilde\beta, H)$ with respect to additive $H$. Even more
difficulties in understanding thermal equilibrium would arise when
one naively considers two systems with the same average $\beta$
and different relative variances, $q_1-1$ and $q_2-1$.

In order to clarify the physical content of the problem of
fluctuating $\tilde\beta$ we turn to the time scale hierarchy
emerging from the microscopic consideration of a dilute gas and
weakly inhomogeneous systems of a finite characteristic size.

\section{The time scale hierarchy}\label{Sec:Hierarchy}

Internal characteristics of a dilute gas with density $n=N/V$ are
known to be the interaction range $r_{\mathrm{int}}$, the mean
distance between particles, $r_{\mathrm{ave}} \simeq n^{-1/3}$,
and the free-flight length $r_{\mathrm{fli}} \simeq
1/(nr_{\mathrm{int}}^2)$. The main parameter of a dilute gas is
the density parameter,
\be
\varepsilon \equiv nr_{\mathrm{int}}^3\ll 1.
\ee
For example, ambient air is characterized by the typical value
$\varepsilon \simeq 10^{-4}$. For small $\varepsilon$ we have
scale hierarchy,
\be
r_{\mathrm{int}} \ll r_{\mathrm{ave}} \ll r_{\mathrm{fli}},
\ee
with $r_{\mathrm{int}}^3:r_{\mathrm{ave}}^3:r_{\mathrm{fli}}^3 =
1:\varepsilon^{-1}:\varepsilon^{-3}$, from which it follows the
time scale hierarchy,
\be
\tau_{\mathrm{int}} \ll \tau_{\mathrm{ave}}
\ll \tau_{\mathrm{fli}},
\ee
where $\tau_{\mathrm{int}} = r_{\mathrm{int}}/v$ is the
interaction time, $\tau_{\mathrm{ave}} = r_{\mathrm{ave}}/v$ is
the average time, $\tau_{\mathrm{fli}} = r_{\mathrm{fli}}/v$ is
the free-flight time, and $v$ is some characteristic velocity of
particles. Typically, for dilute gases $\tau_{\mathrm{int}} \simeq
10^{-12}$ sec and $\tau_{\mathrm{fli}} \simeq 10^{-8}$ sec.

For such systems the quasi-closeness and quasi-equilibrium of
subsystems are well defined, and a description in terms of one-
and two-particle probability distribution functions, or mean local
densities (particle number, momentum, and energy), is rigorously
justified.

One-particle distribution function varies much slower than the
higher order ones due to the smallness of $\varepsilon$, providing
thus quasi-conservation of the mean local densities. At the
microscopic level a quasi-additivity of the (quasi-)local energy
density (hence, of the energy) can be seen from the last term in
its definition,
\be
h(\vec r) = \sum_i \delta(\vec r-\vec r_i)[{\vec p}_i^{\ 2}/2m +
\sum_{j\not=i}\Phi(|\vec r_i-\vec r_j|)].
\ee
Here the last sum makes the biggest contribution for the
coordinates $\vec r_j$ of particles which are at distances less or
about $r_{\mathrm{int}}$ from the point $\vec r$. In fact, whether
$h(\vec r)$ can be taken additive or non-additive depends on the
scale.

We start by considering a sequence of systems,
\be
{\cal N}_1 \subset {\cal N}_2 \subset \cdots \subset {\cal N}_i
\subset \cdots \subset {\cal N}_{\infty},
\ee
generated by the relations between the numbers of particles in
these systems, $1\ll N_1 \ll N_2\ll \cdots
\ll N_i \ll\cdots \ll N_\infty=\infty$. The systems ${\cal
N}_1$ and ${\cal N}_{\infty}$ can be viewed as a smallest system
and an ideal thermostat, respectively, in the equilibrium state.
More specifically, the systems ${\cal N}_i$ are treated as
thermodynamical systems of {\sl characteristic sizes} $r_i$, which
are {\sl external} characteristics, such that
\begin{eqnarray}
r_{\mathrm{int}} \ll r_{\mathrm{ave}}
\ll r_1 
\ll \cdots
\ll r_{k-1}\ll r_{\mathrm{fli}} \ll \cdots \nonumber\\
\cdots \ll r_k \ll \cdots \ll r_{\infty}=\infty.
\end{eqnarray}
Here, we have used the fact that for a dilute gas the free-flight
length is much bigger than the interaction range,
$r_{\mathrm{fli}} \gg r_{\mathrm{int}}$, so that some number
$(k-1)$ of systems fall into the range between $r_{\mathrm{ave}}$
and $r_{\mathrm{fli}}$. In this paper, we will not consider such
systems which require a separate study.

Actually we assume that ${\cal N}_{\infty}$ can be represented as
a collection of weakly interacting {equilibrium} systems ${\cal
N}_1$ so that all bigger systems ${\cal N}_{i}$, $i>1$, are in
{\sl incomplete} equilibrium state at some time scale.

We remind that in the framework of the canonical Gibbs
distribution, the interaction between different ${\cal N}_{1}$'s
is taken negligible, $\xi \equiv H_{\mathrm{int}}/H_{N_1}=0$ (zero
order approximation in small $\xi$). However, the relaxation of
some system ${\cal N}_{i}$, $i>1$, to an equilibrium state is
solely related to $H_{\mathrm{int}}$ so that at least first order
approximation in small $\xi$ should be used to describe the
relaxation process.

The kinetic relaxation length $r_{\mathrm{rel}} \sim
r_{\mathrm{fli}}$, and the associated kinetic relaxation time,
$\tau_{\mathrm{rel}} \sim \tau_{\mathrm{fli}}$, are much bigger
than the correlation length, $r_{\mathrm{cor}} \sim
r_{\mathrm{int}}$ and the associated correlation time,
$\tau_{\mathrm{cor}} \sim \tau_{\mathrm{int}}$, for $\varepsilon
\ll 1$.

In terms of small parameter $\varepsilon$, one can define (see
e.g.~\cite{Klimontovich})
\be
 r_1 = \sqrt{\varepsilon}r_{\mathrm{fli}}, \quad
 \tau_1 = \sqrt{\varepsilon}\tau_{\mathrm{fli}}, \quad
 N_1 = 1/\sqrt{\varepsilon},
\ee
for the {\sl smallest} system ${\cal N}_1$. The size $r_1 =
v\tau_1$ is taken to provide conservation of the local densities
of particle number, momenta and energy in the approximation of
weak inhomogeneity,
\be\l{inhomogeneity}
r_{\mathrm{cor}}/r_1 \ll 1,
\ee
at the time scale $\tau_1$. For $\varepsilon =10^{-4}$, we obtain
the minimal number of particles $N_1=100$. To specify the
hierarchy, we can put $N_i = 1/\varepsilon^{i/2}$.

To each characteristic size $r_i$, $i>k$, we can associate the gas
dynamical relaxation time,
\be\l{gasrel}
\tau_i = \tau_1 (r_i/r_{\mathrm{fli}})^2, \quad i>1,
\ee
so that the time scale hierarchy is
\be
\tau_{\mathrm{int}} \ll \tau_{\mathrm{ave}}
\ll \tau_{\mathrm{fli}} \ll
\tau_k \ll \tau_{k+1} \cdots \ll \tau_{\infty}=\infty.
\ee
For $i \geq k$, the sizes of the systems ${\cal N}_{i}$ are much
bigger than $r_{\mathrm{fli}}$ so that for such systems one more
small parameter can be introduced, $K_i \equiv
r_{\mathrm{fli}}/r_i$, Knudtsen parameter, in terms of which
$\tau_i = \tau_1/K_i^2$.

It should be emphasized that due to $N_1 = 1/\sqrt{\varepsilon}$
the $\varepsilon=0$ approximation in the (nonfluctuating)
canonical Gibbs distribution implicitly means that $N_1 =\infty$
(the approximation of a continuous medium for $N_1$ particle
system). Evidently, the consideration of bigger systems ${\cal
N}_{i}$, $i>1$, becomes of no sense in the $\varepsilon=0$
approximation. This is the point where a large scale dependence is
totally ignored within the approximation of nonfluctuating
distribution function. To keep this dependence one should derive
the quasi-equilibrium distribution function in at least the first
order approximation in $\varepsilon$ and allow distribution
function to fluctuate.

\section{Fluctuating probability distribution functions}

Now we turn to fluctuations. Following to Onsager principle,
kinetics of a linear non-equilibrium thermodynamical system (its
relaxation to an equilibrium state) defines fluctuations in the
system. In turn, the fluctuations of the thermodynamical entities
(evolution of the time correlation functions of their deviations
from equilibrium values) can be equivalently described in terms of
Langevin equation with $\delta$-correlated stochastic sources.
Thus the fluctuations of equilibrium systems can be viewed as a
stationary process with the time scales defined by the relaxation
times $\tau_i$, $i>k$, of the gas dynamical stage of the
relaxation. As shown in Sec.~\ref{Sec:Hierarchy} these time scales
are known to be much bigger than that of the kinetic relaxation
$\tau_{\mathrm{rel}}$ thus providing a local equilibrium
framework.

We are interested in a finite thermostat so that we take some
finite size equilibrium system  ${\cal N}_j$ serving as a
quasithermostat for the system ${\cal N}_i$, i.e., we assume $N_j
\gg N_i$. In accord to the above consideration, the time scale
$\tau_j$ of the quasithermostat ${\cal N}_j$ at which its $\beta$
(kinetic energy per particle) fluctuates is much bigger than the
time scale $\tau_i$ at which any thermodynamical entity of the
system ${\cal N}_i$ fluctuates. In general, there is finite number
of big systems ${\cal N}_{i+1}$, ${\cal N}_{i+2}$,...,${\cal
N}_{j}$, each making contribution to fluctuation of the
temperature with {\sl different} relative variances at
corresponding time scales $\tau_{i+1}$,...,$\tau_{j}$. Observed
marginal distribution for ${\cal N}_i$ will depend on the
observation time $\tau_{\mathrm{obs}}$ as compared to these time
scales. We remark that the observed relative variance of
fluctuating $\beta$ is thus depends on the scale and
$\tau_{\mathrm{obs}}$.

At the time scale $\tau_{\mathrm{obs}} \sim \tau_i$ the canonical
Gibbs distribution makes a very good description of ${\cal N}_i$,
with $\beta$ treated as a constant, and an additive character of
the entropy and mean energy.

At long time scale, $\tau_{\mathrm{obs}} \geq \tau_j \gg \tau_i$,
the fluctuations of $\beta$ can not be ignored so that
Tsallis-like distribution emerges as an effective description,
i.e., we should replace $\beta \to \tilde\beta$, and make the
averaging $\langle e^{-\tilde\beta H_i}\rangle_\beta$, with $\beta
=\langle\tilde\beta\rangle$ and some relative variance. This
results in a non-additive character of the entropy and mean energy
of ${\cal N}_i$. Note however that in the latter case the usual
quasi-equilibrium (thermal equilibrium) still holds for any given
value of $\tilde\beta$ at the small time scale $\tau_i$. This
picture meets that assumed by Beck~\cite{Beck}.

For $j\to\infty$ (an ideal thermostat), we have $\tau_j\to\infty$
so that $\beta$ is constant at any time scale $\tau > \tau_i$,
thus $q=1$ for ${\cal N}_\infty$.

To put the above consideration on solid grounds, below we outline
statistical mechanics approach to fluctuations, within a dynamical
framework.

We can treat the quasithermostat ${\cal N}_j$ as an equilibrium
system, which can be described in pure additive thermodynamical
terms, with the canonical Gibbs distribution,
$Z^{-1}\exp[-H_j/kT]$, providing the averaging in ${\cal N}_j$.

However, kinetic and thermodynamic fluctuations naturally occur in
the quasi-equilibrium system ${\cal N}_j$ and can be calculated in
a {\sl linear} approximation (small fluctuations) for the case of
weak inhomogeneity of the system, Eq.~(\ref{inhomogeneity}). We
notice that rigorous description of fluctuations is out of scope
of the canonical Gibbs distribution and the one-particle
distribution function entering Boltzmann equation.

In a dynamical framework, one should consider Bogolyubov equations
with a short-range interaction potential, $\Phi(|{\vec r}_m-{\vec
r}_n|)$, for point-like particles (providing convergence of the
integral for local energy density), small density parameter
$\varepsilon$ (i.e., three- and higher order particle collisions
are ignored), and weak inhomogeneity (providing small gradient of
the one-particle distribution function), and account for the
two-particle correlation function with {\sl a partial relaxation}
of initial correlations. This implies fluctuations of the
one-particle distribution function; see e.g. \cite{Klimontovich}
and review \cite{Gantsevich}. The form of the resulting kinetic
equations allows one to study separately implications related to
Brownian effect (a discrete collision structure of the medium)
thus discarding influence of fluctuations on kinetic processes.

The resulting picture is equivalent to consideration of some
Langevin equations with $\delta$-correlated stochastic sources,
from which one can extract two-point correlation functions and
spectral densities of the gas density, velocity, and pressure, for
a local equilibrium state. In the local equilibrium approximation
correlation functions of the stochastic sources can be found also
by phenomenological methods of non-equilibrium
thermodynamics~\cite{Landau}.

The spectral densities can be used to estimate the gas dynamical
relaxation time, $1/\tau \sim \nu {\vec k}^2 \sim \kappa {\vec
k}^2$, where $\nu$ is kinematic viscosity, $\kappa$ is heat
conductivity, and ${\vec k}$ is wave vector. For a finite system
of characteristic size $L$, the wave vector ${\vec k}$ is
naturally bounded from below, $|{\vec k}| \geq 1/L$, from which
with the use of estimation $\nu\sim\kappa\sim vr_{\mathrm{fli}}$
it follows Eq.~(\ref{gasrel}).

We focus on the temperature fluctuations, for which one can obtain
the approximate variance,
\be\l{DeltaT}
\langle(\Delta\tilde
T)^2\rangle \simeq kT^2 (dT/dE)_{V},
\ee
with temperature $\tilde T$ being normally distributed (in this
approximation), and $T$ in the r.h.s. is given in terms of
equilibrium value. This means that $\tilde T$ fluctuates about the
average $T$ with the relative variance $k/C_v$, where $C_v =
(dE/dT)_{V} \sim N_j$ is a heat capacity of the quasithermostat
${\cal N}_j$, thus ending up with the consideration of
Sec.~\ref{Sec:Fluctuations}.

It is worthwhile to note that the entropy $S$ of ${\cal N}_j$
fluctuates as well, $\langle(\Delta\tilde S)^2\rangle \simeq
kC_p$, $C_p \sim N_j$.

Clearly, natural fluctuations occur not only in the
quasithermostat ${\cal N}_j$ but also in the considered smaller
system ${\cal N}_i$. However, we can ignore these fluctuations as
they make small contribution to specific fluctuation of the
temperature $\tilde T$ governed by the quasithermostat, due to
$N_i \ll N_j$.

We point out that a major difference from the above consideration
arises when turning to fully developed turbulence considered in
\cite{Beck}. Beck used {\sl statistical} temperature related to
kinetic coefficients (fluctuations of which are evidently not
related to a finiteness of quasithermostat) and have averaged the
probability distribution function (describing a stationary
non-equilibrium state) derived from a nonlinear Langevin equation
which is different from the canonical Gibbs distribution.
Fluctuations of the friction coefficient (or viscosity) are seem
to be of a kinetic type originating from large scale fluctuations
of the phase-space density function averaged over the smallest
scale. Such fluctuations may have a big relative variance as they
are not related to the particle number of some big system, with
the main contribution to the energy dissipation happening at the
smallest scale being fluctuations, $\tilde u_i \sim \delta u_i$,
where $\tilde u_i$ is velocity and $\delta u_i$ is fluctuation of
the velocity. Indeed, for Kolmogorov (the smallest) length scale
$\eta$ of the turbulent flow Beck has derived under simple
assumptions a theoretical value of the entropic parameter,
$q=3/2$, which essentially deviates from unity (the relative
variance is considerable, $q-1=1/2$). One can see \cite{Beck} that
Tsallis formalism does work in this case as well. For bigger
scales, fluctuations of the energy dissipation decrease so that in
zero order approximation in $\eta/r$ the energy dissipation is
constant. Clearly, one should assign the value $q=1$ for this
large scale asymptotics.

Fluctuations of $\tilde\beta$ in the canonical Gibbs distribution
mean that we deal with a {\sl fluctuating distribution function}.
The $\chi^2$ {\sl average} value of this fluctuating distribution
function is Tsallis distribution,
\be
\langle e^{-\tilde\beta H} \rangle_{\beta} = e_{q-1}^{-\beta H}.
\ee
The relative variance of this fluctuation is
\be
\frac{\langle e^{-2\tilde\beta H} \rangle_{\beta} -\langle
e^{-\tilde\beta H} \rangle_{\beta}^2}{\langle e^{-\tilde\beta H}
\rangle_{\beta}^2} = s\ln_{s}\left(1+
\frac{s^2\beta^2H^2}{1+2s\beta H}\right),
\ee
where $s=1-q$. For small $(1-q)$, this relative variance behaves
as $(1-q)^3$. Also, up to a first order in $(q-1)$, we obtain
\be\l{lin}
\exp_{q-1}[-\beta H] \simeq \left(1+ \frac{q-1}{2}\beta^2
H^2\right)\exp[-\beta H].
\ee
It is remarkable to note that this function is non-factorizable
with respect to $H$ because of the presence of $H^2$.

 From the above dynamical consideration it follows that
fluctuations of a probability distribution function arise
naturally in studying both equilibrium and non-equilibrium
phenomena. However, in studying equilibrium state one usually uses
non-fluctuating distribution function, thus discarding specific
effects such as wellknown Brownian one. Namely, the
$\varepsilon=0$ approximation means a continuous medium
approximation, at which Brownian motion can not be described, even
for the smallest physical equilibrium system ${\cal N}_1$
consisting of, e.g., $N_1=100$ particles for
$\varepsilon=10^{-4}$.

In contrast, the use of a phenomenological Tsallis approach makes
possible to account for some of the effects related to natural
fluctuations of an equilibrium distribution function or that
corresponding to a stationary non-equilibrium process.

Finally, we point out that even {\sl small} fluctuations of the
one-particle distribution function, $\delta f/f \ll 1$, make an
essential contribution (as big as Boltzmann integral which is
treated as a leading term) to Boltzmann equation with Langevin
sources \cite{Klimontovich}. This indicates importance of
accounting for fluctuations of probability distribution functions,
both for equilibrium and non-equilibrium states, at some scales.

\section{Discussion}\label{Sec:Discussion}

The microcanonical approach in the form presented in this paper
evidently can not cover a universal character of Tsallis
formalism. First, since we assume $M\gg 1$ the parameter $q-1$ is
very close to zero that allows us to use expansion in $q-1$ and
calculate small corrections to the canonical Gibbs distribution,
while for considerable $q-1$ the arguments of our derivation in
Sec.~\ref{Sec:Quasi} break down. Second, we have ignored the
interaction energy $H_{\mathrm{int}}$ that is not valid for
systems with long-range microscopic interactions to which Tsallis
statistics is applied. Also, the nonextensivity can arise not only
as the effect of a finite heat bath. For example, Tsallis
distribution arises in the description of polydispersed colloids
with the fluctuating parameter being particle size \cite{Johal},
and was used to describe statistical properties of fully developed
turbulence \cite{Beck2, Wilk}; see also review in \cite{Tsallis3}.

In general, fluctuations of the parameter $\tilde\beta$ (or
statistical temperature) and averaging of the fluctuating
distribution function over $\chi^2$-like distribution can be
viewed as a defining feature of Tsallis-like nonextensive
statistical distributions, regardless to the origin of
fluctuations.

Adopting this viewpoint, one can try various types of distribution
of $\tilde\beta$ in replace of $\chi^2$ distribution, and allow
considerable relative variance of $\tilde\beta$. For example, it
is natural to assume Poisson distribution for the random momenta
$\vec p_i$ in replace of the normal distribution; see also
\cite{Sattin}. Also, one may be interested in a more detailed
description and account for higher order momenta such as
biquadratic deviations of~$\tilde\beta$.

From this point of view, Tsallis formalism can be treated as a
suitable universal tool to deal with fluctuating distribution
functions without turning to underlying kinetic issues. However,
it should be noted that the use of an {\sl averaged} fluctuating
distribution function means {\sl incomplete} statistical
description.

Finally, we stress that the very notions of extensivity
(additivity) and intensivity in thermodynamics are essentially
based on the requirement that the system is homogeneous, which is
provided for big systems with weak interactions or, more
precisely, in the thermodynamic limit, $N, V \to \infty$, $N/V =$
const. These notions make no strict sense for inhomogeneous
systems such as finite systems or systems characterized by the
size of the order of correlation length $r_{\mathrm{cor}}$. As a
consequence, in this case there is no thermodynamical equivalence
between equilibrium microcanonical and canonical ensembles.


\begin{thebibliography}{12}


\bibitem{Tsallis}
 C. Tsallis, J. Stat. Phys. {\bf 52}, 479 (1988).

\bibitem{Tsallis2}
 C. Tsallis, R.S. Mendes, and A.R. Plastino, Physica {\bf 261A},
 534 (1998).

\bibitem{Tsallis3}
 C. Tsallis, {\it Entropic nonextensivity: a possible measure of
 complexity}, cond-mat/0010150.

\bibitem{Abe}
 S. Abe, Phys. Lett. {\bf 271A}, 74 (2000);
 S. Abe, S. Martinez, F. Pennini, A. Plastino, Phys. Lett. {\bf 281A}, 126 (2001);
 S. Abe, Phys. Rev. {\bf 63E}, 061105 (2001);
 S. Abe and A.K. Rajagopal, {\it Nonuniqueness of canonical ensemble theory arising from
microcanonical basis}, quant-ph/9911097;
 {\it Nonadditive conditional entropy and its significance for local
realism}, quant-ph/0001085;
 {\it Justification of power-law canonical distributions based on
generalized central limit theorem}, cond-mat/0003380;
 A.K. Rajagopal and S. Abe, Phys. Rev. Lett. {\bf 83}, 1711 (1999);
 {\it Statistical mechanical foundations for systems with
nonexponential distributions}, cond-mat/0003493.

\bibitem{Beck}
 C. Beck, {\it Generalized statistical mechanics and fully developed
turbulence}, cond-mat/0110073.

\bibitem{Johal}
R. Johal, {\it An interpretation of Tsallis statistics based on
polydispersity}, cond-mat/9909389.

\bibitem{Beck2}
 C. Beck, Physica {\bf 277A}, 115 (2000);
 Phys. Lett. {\bf 287A}, 240 (2001);
{\it Dynamical foundations of nonextensive statistical mechanics},
to appear in Phys. Rev. Lett. (2001);
 {\it Non-additivity of Tsallis entropies and fluctuations of temperature},
cond-mat/0105371.

\bibitem{Wilk}
G. Wilk and Z. Wlodarczyk, Phys. Rev. Lett. {\bf 84}, 2770 (2000).

\bibitem{Almeida}
M. P. Almeida, {\it Generalized entropies from first principles},
cond-mat/0102112.

\bibitem{Plastino}
A. Plastino; A. R. Plastino,
Brazilian J. of Phys. {\bf 29} , no. 1 (1999).

\bibitem{Aringazin}
 A.K. Aringazin, Phys. Lett. {\bf 314B}, 333 (1993);
 A.K. Aringazin, V.V. Arkhipov, and A.S. Kudusov, {\it BRST approach to
Hamiltonian systems}, hep-th/9811026.


\bibitem{Abe2}
S. Abe and A.K. Rajagopal, {\it Macroscopic thermodynamics of
equilibrium characterized by power-law canonical distributions},
cond-mat/0009400.

\bibitem{Klimontovich}
Yu.L. Klimontovich, {\it Statistical Physics} (Nauka, Moscow,
1982) (in Russian).

\bibitem{Gantsevich}
S.V. Gantsevich, V.L. Gurevich, R. Katilius,
Rev. Nuovo Cimento {\bf 2}, 1 (1979).

\bibitem{Landau}
E.M. Lifshitz and L.P. Pitaevski, {\it Statistical Physics},
Part~2. Nauka, Moscow, 1978. (in Russian).

\bibitem{Sattin}
F. Sattin and L. Salasnich, {\it Multi-parameter generalization of
nonextensive statistical mechanics}, physics/0110055.




\end{thebibliography}
\end{document}